\documentclass[twocolumn,english,prl,amssymb,aps,superscriptaddress,showpacs,twocolumn,amsmath,showkeys,floatfix]{revtex4-1}
\usepackage[T1]{fontenc}
\usepackage[latin9]{inputenc}
\usepackage[active]{srcltx}
\usepackage{textcomp}
\usepackage{amsmath}
\usepackage{graphicx}
\usepackage{color}
\usepackage{esint}
\usepackage{caption}

\makeatletter
\usepackage{babel}

\makeatother

\begin{document}

\title {Investigation of the noise figure in degenerate dual-pump phase sensitive amplifier using a multi-wave model}
\author{Yousra Bouasria}
\affiliation{Equipe Sciences de la Mati\`ere et du Rayonnement (ESMaR), Faculty of Sciences, Mohammed V University, Rabat, Morocco}
\author{Debanuj Chatterjee}
\affiliation{Universit\'e Paris-Saclay, CNRS, ENS Paris-Saclay, CentraleSup\'elec, LuMIn, Gif-sur-Yvette, France}
\author{Weilin Xie}
\affiliation{School of Optics and Photonics, Beijing Institute of Technology, 100081 Beijing, China}
\author{Ihsan Fsaifes}
\affiliation{Ecole Polytechnique, Institut Polytechnique de Paris, 91128 Palaiseau Cedex, France}
\author{Fabienne Goldfarb}
\affiliation{Universit\'e Paris-Saclay, CNRS, ENS Paris-Saclay, CentraleSup\'elec, LuMIn, Gif-sur-Yvette, France}
\author{Yassine Hassouni}
\affiliation{Equipe Sciences de la Mati\`ere et du Rayonnement (ESMaR), Faculty of Sciences, Mohammed V University, Rabat, Morocco}
\author{Fabien Bretenaker}
\affiliation{Universit\'e Paris-Saclay, CNRS, ENS Paris-Saclay, CentraleSup\'elec, LuMIn, Gif-sur-Yvette, France}
\affiliation{Light and Matter Physics Group, Raman Research Institute, Bangalore 560080, India}

\begin{abstract} 
A semi-classical 7-wave model is developed to investigate the noise performances of a degenerate dual-pump phase sensitive amplifier. This approach takes into account the transfer to the signal,  through multiple four-wave mixing processes, of the vacuum fluctuations injected in the high-order waves. This effect leads to a degradation of the noise figure of the amplifier with respect to the 0 dB value predicted by the usual 3-wave model. However, it is proved that a careful choice of the fiber dispersion allows to use the high-order waves to enhance the signal gain without degrading the noise figure above 1 dB.
\end{abstract}


\maketitle
\section{Introduction}
Phase-sensitive fiber-optic parametric amplifiers (PS-FOPAs) \cite{tong2011towards}, relying on four-wave mixing (FWM) \cite{marhic2008fiber} in highly nonlinear fibers (HNLFs), have attracted a lot of attention due to their broad gain spectrum \cite{hansryd2002fiber}, their potentially noiseless amplification capability \cite{levenson1993reduction}, and their compatibility with fiber communication systems \cite{tong2013low}. Such optical amplifiers are capable of amplifying a shot-noise limited input signal without degrading the signal-to-noise (SNR) ratio, and can thus exhibit a quantum limited noise figure (NF) of 0 dB, much smaller than the 3 dB limit of conventional phase insensitive amplifiers (PIAs) \cite{caves1982quantum,yamamoto2003noise}. This unique capability of phase sensitive amplifiers (PSAs) has led to recent demonstrations of phase and amplitude regeneration of complex encoded signals \cite{slavik2010all}, noise reduction \cite{karlsson2015transmission}, and mitigation of nonlinear phase impairments \cite{olsson2015nonlinear}.

Different types of theoretical approaches have been developed to predict the noise performances of PSAs. Some of them rely on a semi-classical treatment of the optical fields \cite{olsson1989lightwave,donati1997noise,tong2010noise,lundstrom2012phase,tong2012ultralow}, some others on a fully quantum mechanical treatment \cite{mckinstrie2006quantum,mckinstrie2010field,marhic2012noise,marhic2013quantum,inoue2016quantum}. In the current literature, most of these works have been based on 3-wave degenerate FWM or 4-wave non-degenerate FWM schemes. However, when the nonlinearity becomes large, new frequencies are created by extra FWM processes,  thus making the 3- or 4-wave approach irrelevant. For example, in the case of the dual-pump configuration with degenerate signal and idler, some situations require the development of a 7-wave model involving 22 FWM processes occurring simultaneously along the fiber \cite{xie2015investigation, baillot2016phase}. Some further generalizations have even considered up to 27 modes \cite{Qian2017}. In the case of the 7-wave model, the fact that more than three waves are involved in the process has led to the prediction of possible gain enhancements \cite{xie2015investigation} and applications to signal regeneration \cite{xie2017optimization}, which are very attractive for applications. The question then arises to know whether such gain enhancements are accompanied with enhancements of the PSA noise or not.

The calculation of the NF in such a multi-wave situation is a topic of active investigation. Several attempts have been initiated \cite{mckinstrie2004quantum,mckinstrie2005quantum}, but until recently none of them provided a complete theoretical description of the signal NF when multiple waves are accounted for.
Recently, a remarkable quantum mechanical approach has been developed to calculate the noise performances of a degenerate dual-pump PSA in the framework of a 7-wave model  \cite{inoue2019influence}. This study reported the influence  on the PSA NF of some of the FWM processes among the ones involved in the interaction between the seven waves. However, according to the phase matching condition \cite{agrawal2000nonlinear}, the 7-wave model may involve up to 22 FWM processes occurring simultaneously along the fiber \cite{xie2015investigation, baillot2016phase}. We can thus expect some situations to occur, in which  all the possible mechanisms should be taken into account in the NF calculation, in order to get more accurate results. 

Therefore, our aim here is to investigate the impact of high-order waves on the noise limit of a 7-wave PSA, in terms of phase matching condition and pump wavelength allocation. Since it is difficult to theoretically assess  the influence of all the 22 FWM processes following the quantum mechanical procedure of Ref.\,\cite{inoue2019influence}, we choose a semi-classical approach and rely on numerical simulations along those of Ref.\,\cite{xie2015investigation}. The method investigated here takes into account all the 22 FWM processes occurring among the considered 7 waves, contrary to the quantum model presented in Ref.\,\cite{inoue2019influence}, which restricts to only some of them. Furthermore, we calculate here the amount of vacuum noise transferred from each of the empty input modes to the signal, thus predicting which of the input waves mostly contributed to the degradation of the PSA NF for the signal. 

The paper is organized as follows: In Section \ref{Sec2}, we compare the gain spectra predicted by different approaches in the domains of parameters we are interested in, in order to choose which approach is better adapted to our situation. Then, Section \ref{Sec3} is devoted to the semi-classical derivation of the noise figure in the framework of our 7-wave semi-numerical approach. Finally, Section \ref{Sec4} applies this model to different situations in order to determine how strongly the presence of high-order waves is detrimental to the PSA noise figure, and we discuss the underlying physical mechanisms.

\section{Comparison between different gain models}\label{Sec2}
In this section, we describe the different models used to calculate the gain and the NF of a dual-pump PSA with degenerate signal and idler, and we compare the gains that they predict in different situations.

\subsection{3-wave model with undepleted pumps}\label{Sec2A}
\begin{figure}[h]
\centering
\includegraphics[width=0.5\columnwidth]{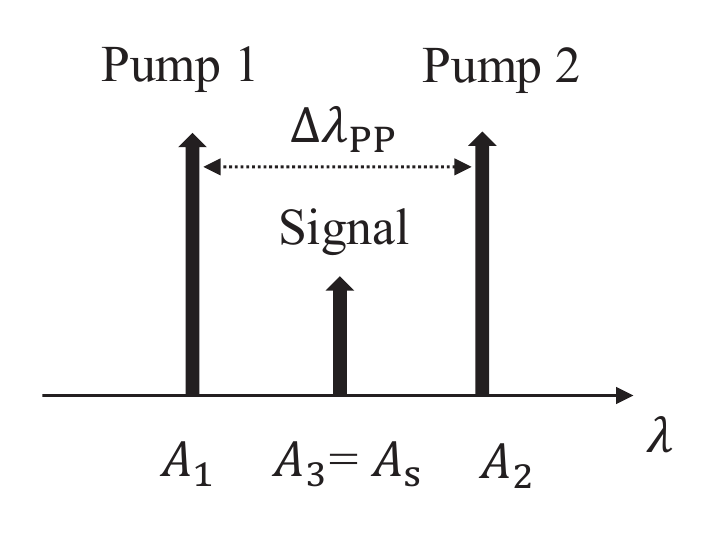}
 \caption{Schematic representation of the PSA in the 3-wave model. $\Delta\lambda_{\mathrm{PP}}$ is the pump-pump wavelength separation.}\label{Figure01}
\end{figure}

We consider in this paper the PSA architecture schematized in Fig.\,\ref{Figure01}, in which a degenerate signal and idler is amplified by FWM interaction with two symmetrically located pumps in a highly nonlinear fiber (HNLF). The complex amplitudes of the electric fields of the three co-polarized waves are labeled $A_j$ with $j=1..3$, as shown in Fig.\,\ref{Figure01}.

\subsubsection{Gain}
In the small signal regime, i. e., when depletion of the pumps  can be neglected, and neglecting  fiber attenuation, the phase sensitive gain for the signal power is given by \cite{agrawal2000nonlinear}:
\begin{multline}\label{Eq6}
    G_{\mathrm{PSA}} = 1 + \Big\{1+\frac{\kappa^{2} + 16 \gamma^{2} P_{1} P_{2} + 8\gamma\kappa\sqrt{P_{1}P_{2}} \cos{\Theta}}{4 g^{2}}\Big\}\\ \times\sinh^{2}{(g L)} \\
    + \frac{2\gamma\sqrt{P_{1}P_{2}} \sin{\Theta}}{g} \sinh{(2gL)}\ ,
\end{multline}
where $\gamma$ is the nonlinear Kerr coefficient of the fiber, $P_{1}$ and $P_{2}$ are the incident pump powers, and $L$ is the fiber length. The relative phase $\Theta$ between the waves is given by:
\begin{equation}\label{Eq7}
    \Theta = 2 \phi_{3}  - \phi_{1} - \phi_{2}\,,
\end{equation}
where $\phi_{j}$  is the input phase of field $A_j$ $(j=1..3)$. The total phase mismatch $\kappa$ is defined as:
 \begin{equation}\label{Eq8}
     \kappa = \Delta\beta + \gamma (P_{1} + P_{2})\ ,
 \end{equation}
 where the first term $\Delta\beta$ represents the linear phase mismatch between the interacting waves and the second term represents the nonlinear phase mismatch, assuming that the pumps power are much stronger than the signal power. Finally, the parametric gain coefficient $g$ is given by:
\begin{equation}\label{Eq9}
    g=\sqrt{(2\gamma)^{2}P_{1}P_{2} - \left(\frac{\kappa}{2}\right)^{2}}\ .
\end{equation}
From Eq.\,(\ref{Eq6}), we see that the gain maximum $G_{\mathrm{max}}$ and minimum $G_{\mathrm{min}}$ are reached  when the relative phase $\Theta$ is equal to $2k\pi$ or $(2k+1)\pi$, respectively, with $k$ an integer \cite{tong2011towards}.

\subsubsection{Noise figure}
The NF is a measure of how much excess noise is added to the signal by the amplifier. It is defined as the ratio between the signal-to-noise ratio (SNR) at the input of the amplifier $SNR_{in}$ and the SNR at the output $SNR_{out}$:
\begin{equation}\label{Eq3}
NF = \frac{SNR_{\mathrm{in}}}{SNR_{\mathrm{out}}}\ .
\end{equation}
The SNR is defined as the ratio of the electrical signal power to the electrical noise power measured using an ideal photo-detector \cite{tong2013low}:
\begin{equation}\label{Eq4}
NF = \frac{\langle I_{\mathrm{s,in}} \rangle ^{2} / \langle \Delta I_{\mathrm{s,in}}^{2} \rangle}{\langle I_{\mathrm{s,out}} \rangle ^{2} / \langle \Delta I_{\mathrm{s,out}}^{2}\rangle}\,,
\end{equation}
Where $\langle I_{\mathrm{s}} \rangle$ denotes the mean photo-current after square law-detection and  $\langle\Delta I_{\mathrm{s}}^2\rangle = \langle \left(I_{\mathrm{s}} - \langle I_{\mathrm{s}} \rangle\right)^2\rangle$ its variance.
%

If the detector is ideal, the NF can be also expressed in terms of light power as:
\begin{equation}\label{Eq5}
\begin{split}
NF & = \frac{P_{\mathrm{s,in}}/P_{\mathrm{noise,in}}}{P_{\mathrm{s,out}}/P_{\mathrm{noise,out}}} \\
 & = \frac{P_{\mathrm{s,in}} (G P_{\mathrm{noise,in}} + P_{\mathrm{noise,extra}})}{P_{\mathrm{noise,in}} (G P_{\mathrm{s,in}})} \\
 & = \frac{G P_{\mathrm{noise,in}} + P_{\mathrm{noise,extra}}}{G P_{\mathrm{noise,in}}}\,,
\end{split}
\end{equation}
where $P_{\mathrm{s,in}}$ and $P_{\mathrm{s,out}}$ are the input and output signal powers, respectively,  $P_{\mathrm{noise,in}}$ and $P_{\mathrm{noise,out}}$ are the input and output signal noise powers, respectively, and  $P_{\mathrm{noise,extra}}$ is the extra noise power induced by the amplifier itself.

For the three-wave model mentioned here, the NF associated with the gain of Eq.\,(\ref{Eq6}) reads \cite{ferrini2014symplectic}:
\begin{equation}\label{Eq14}
NF = \frac{G_{\mathrm{max}} + G_{\mathrm{min}}}{G_{\mathrm{PSA}}}\ .
\end{equation}
In particular, at gain maximum ($\Theta=0$), this expression becomes
\begin{equation}\label{Eq14N1}
NF = 1+\frac{G_{\mathrm{min}} }{G_{\mathrm{max}}}\ ,
\end{equation}
which, in the case of the three-wave model for which $G_{\mathrm{min}} =1/G_{\mathrm{max}}$, becomes
\begin{equation}\label{Eq15}
NF = 1 + \frac{1}{G^{2}_{\mathrm{max}}}\ .
\end{equation}
Thus, in the limit of high gain values $(G_{\mathrm{max}}\gg 1)$, the signal NF for a 3-wave PSA takes the quantum limited value of 1 (0~dB).

\subsection{7-wave numerical model}\label{Sec2B}
\begin{figure}[h]
\centering
\includegraphics[width=0.9\columnwidth]{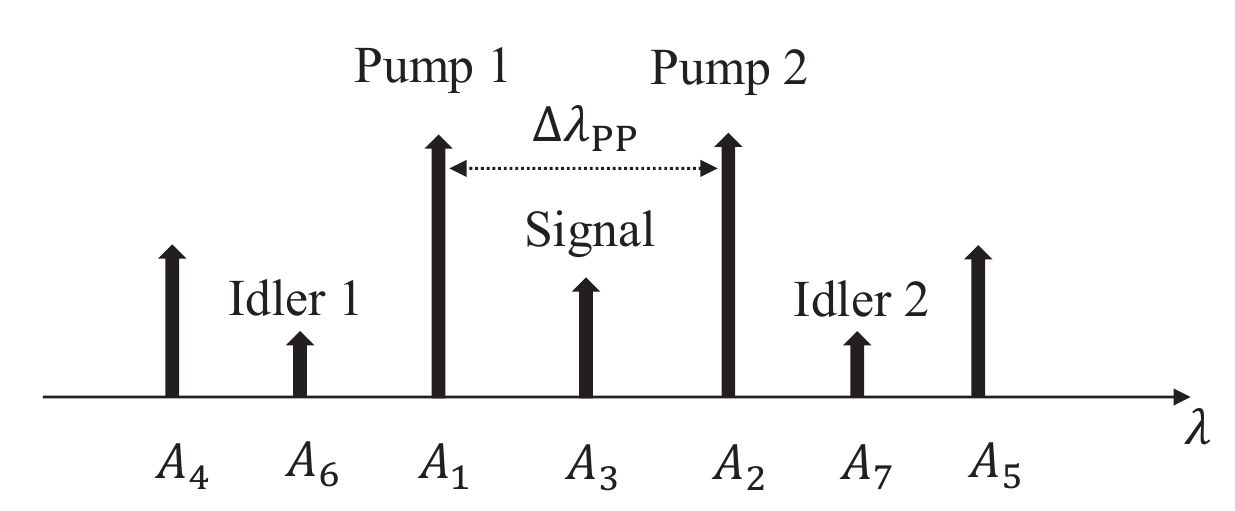}
 \caption{Schematic representation of the 7-wave model. $\Delta\lambda_{\mathrm{PP}}$ is the pump-pump wavelength separation. $A_{j}$ and $\lambda_{j}$ are the complex field amplitude and wavelength of the $j^{th}$ wave.}\label{Figure02}
\end{figure}

Launching two intense pump fields in a HNLF, as shown in Fig.\,\ref{Figure01}, can lead to the creation of many extra tones by cascaded FWM interactions \cite{thompson1991nonlinear,hart1994conservation,trillo1994nonlinear,hart1998dynamical}. Ultimately, such cascaded interactions can lead to the creation of a whole frequency comb \cite{sefler1998frequency,xu2009multiple,myslivets2012generation}. Without reaching such extremities, the 7-wave model considers the case, schematized in Fig.\,\ref{Figure02}, where four more waves are created: two so-called high-order pumps, labeled 4 and 5, which are chiefly generated by FWM between the two incident pumps, and two so-called high-order idlers, labeled 6 and 7, which are mainly generated by FWM with the signal and one of the two pumps.



In this model, the evolution of the amplitudes of the seven waves along the fiber is described by a set of seven coupled equations \cite{baillot2016phase, agrawal2000nonlinear}. For brevity, we give here only the equation regarding the signal wave evolution:
\begin{multline}\label{Eq16}
\frac{dA_{3}}{d z} = -\frac{\alpha}{2} A_{3} + i\gamma \left\{ \Bigg .\left[|A_{3}|^{2} + \sum_{j=1, j\neq 3}^{7} |A_{j}|^{2}\right] A_{3}\right. \\
+ A_{1}A_{1}A^{*}_{6} e^{-i \Delta\beta_{3611}} 
+ A_{2}A_{2}A^{*}_{7} e^{-i \Delta\beta_{3722}} 
+ 2 A_{1}A_{2}A^{*}_{3} e^{-i \Delta\beta_{33   12}}\\
+ 2 A_{1}A_{5}A^{*}_{7} e^{-i \Delta\beta_{3715}}
+ 2 A_{1}A_{6}A^{*}_{4} e^{-i \Delta\beta_{3416}} 
+ 2 A_{1}A_{7}A^{*}_{2} e^{-i \Delta\beta_{3217}}\\
+ 2 A_{2}A_{4}A^{*}_{6} e^{-i \Delta\beta_{3624}} + 2 A_{2}A_{6}A^{*}_{1} e^{i \Delta\beta_{2631}}
+ 2 A_{2}A_{7}A^{*}_{5} e^{i \Delta\beta_{2735}}
\\
+ 2 A_{7}A_{4}A^{*}_{1} e^{i \Delta\beta_{7431}} 
+ 2 A_{4}A_{5}A^{*}_{3} e^{i \Delta\beta_{4533}} + 2 A_{6}A_{7}A^{*}_{3} e^{i \Delta\beta_{6733}} \\
\left.+ 2 A_{5}A_{6}A^{*}_{2} e^{i \Delta\beta_{6532}} \right\}\ .
\end{multline}
Here $\alpha$ is the fiber attenuation coefficient and $\Delta\beta_{mnkl} =\beta(\omega_{m}) + \beta(\omega_{n}) -\beta(\omega_{k}) - \beta(\omega_{l}) $ is the linear phase mismatch of the interacting waves, with $\beta(\omega_{j})$, ($j = 1,..,7$)  the propagation constant at frequency $\omega_{j}$ \cite{agrawal2000nonlinear}.

The first term on the right-hand side of Eq. (\ref{Eq16}) represents the fiber attenuation and the second term gives rise to nonlinear phase-shifts due to self- and cross-phase modulation. The last terms hold for the different FWM mechanisms involving the signal $A_3$ and depend on $\Delta\beta_{mnkl}$, which in turn depends on the fiber dispersion. 

In practice, the efficiency of the FWM process involving the waves $m, n, k,$ and $l$ in Eq. (\ref{Eq16}) is governed by the total phase mismatch:
\begin{equation}\label{Eq17}
\kappa_{mnkl} =  \Delta\beta_{mnkl} +\gamma P_{mnkl} \,,
\end{equation}
where $\gamma P_{mnkl} = \gamma (P_{k} +P_{l} - P_{m} - P_{n})$ represents the nonlinear phase mismatch with $P_{j}$ the power of wave $j$. 


The 7-wave model involves 22 FWM processes, divided into 13 non-degenerate  and 9 degenerate processes. Numerical methods are used to solve this model and extract for example the signal gain spectrum \cite{xie2015investigation, baillot2016phase}.

\subsection{7-wave semi-quantum model}\label{Sec2C}
Recently, Inoue introduced a new approach \cite{inoue2019influence} to calculate the NF of a dual-pump PSA, taking into account the existence of the 7 waves of Fig.\,\ref{Figure02}. This very nice approach is based on two sets of propagation equations. In the first set, the coupled evolution of the pumps labeled 1 and 2 and the so-called high-order pumps labeled 4 and 5 is treated \emph{classically}: 
\begin{eqnarray}
\hspace{-0.3cm}\frac{dA_{1}}{d z}& = &i\gamma \left[ \left(|A_{1}|^{2}+2 |A_{2}|^{2}\right)A_{1}\right.\nonumber\\
&&\left.+ 2 A_{4}A_{2}A^{*}_{1} e^{-i \Delta\beta_{2411}} + A_{2}A_{2}A^{*}_{5} e^{i \Delta\beta_{1522}}\right]\,,\label{eq12}\\
\frac{dA_{2}}{d z}& =& i\gamma \left[\left(2 |A_{1}|^{2}+ |A_{2}|^{2}\right)A_{2}\right.\nonumber\\
&&\left.+ 2 A_{5}A_{1}A^{*}_{2} e^{-i \Delta\beta_{1522}} + A_{1}A_{1}A^{*}_{4} e^{i\Delta\beta_{2411}}\right]\,,\label{eq13}\\
\frac{dA_{4}}{d z}& = &i\gamma \left[ \Big(2|A_{1}|^{2}+2 |A_{2}|^{2}\Big)A_{4} + A_{1}A_{1}A_{2}^{*} e^{i \Delta\beta_{2411}}\right]\,,\label{eq14}\\
\frac{dA_{5}}{d z}& = &i\gamma \left[\Big(2|A_{1}|^{2}+2 |A_{2}|^{2}\Big)A_{5} + A_{2}A_{2}A_{1}^{*} e^{i \Delta\beta_{1522}}\right]\,.\label{eq15}
\end{eqnarray}
Equations (\ref{eq12}-\ref{eq15}) permit to take into account the depletion of the pumps induced by the creation of fields 4 and 5, resulting in a calculation of the evolution of the powers of the pumps along the fiber.

Second, the result obtained for the evolution of the pump powers is used as an input in another set of equations describing the evolution of the \emph{quantum} signal field labeled 3 coupled to the quantum fields labeled 6 and 7, namely: 
\begin{eqnarray}\label{Eq c2}
\frac{d\hat{a}_{3}}{d z}& =& 2 i\gamma \hat{a}_{1}\hat{a}_{2}\hat{a}^{\dagger}_{3} e^{-i \Delta\beta_{3312}} 
+ i\gamma \hat{a}_{1}\hat{a}_{1}\hat{a}^{\dagger}_{6} e^{-i \Delta\beta_{3611}}\nonumber \\
&&+  i\gamma \hat{a}_{2}\hat{a}_{2}\hat{a}^{\dagger}_{7} e^{-i \Delta\beta_{3722}}\;,\\
\frac{d\hat{a}_{6}}{d z} &= &i\gamma \hat{a}_{1}\hat{a}_{1}\hat{a}_{3}^{\dagger} e^{-i \Delta\beta_{3611}}\;,\\
\frac{d\hat{a}_{7}}{d z}& =& i\gamma \hat{a}_{2}\hat{a}_{2}\hat{a}_{3}^{\dagger} e^{-i \Delta\beta_{3722}}\;.
\end{eqnarray}
The approach of Ref.\,\cite{inoue2019influence} relies on dividing the fiber length into small segments in which the pump powers are supposed to be constant, leading to the derivation of transfer matrices between the input and output operators for fields 3, 6, and 7 for each segment. Finally, the total transfer matrix of the whole fiber length is obtained by multiplying all the transfer matrices from the fiber input to the output ends.
%

In summary, this model takes into account, on the one hand, saturation of the gain due to the high-order pumps 4 and 5, but not by the signal 3 and the high-order idlers 6 and 7. On the other hand, it considers the transfer of vacuum fluctuations from fields 6 and 7 to the signal, using only 3 processes out of the 13 processes of Eq.\,(\ref{Eq16}).

\subsection{Comparison between the models}
It has been clearly established that the three-wave model of Section \ref{Sec2A} is not sufficient to give an accurate description of the PSA gain when the frequency separation between the pumps is small and/or the pump powers are large, because it neglects the creation of the extra waves shown in Fig.\,\ref{Figure02} \cite{xie2015investigation}. The  intermediate model of Section \ref{Sec2C} is thus an  attractive alternative. But before using it to calculate the PSA NF, we check whether the range of parameters lies within the domain of applicability of this model. To this aim, we compare the gain spectra obtained from the three models for two different sets of parameters.
\begin{table}[h]
\centering
\begin{tabular}{l c c}
\hline
Parameters &  (a)  &  (b) \\
\hline
$P_{1,in}$   & 0.2\,W & 0.1\,W \\
$P_{2,in}$  & 0.2\,W & 0.1\,W \\
$P_{\mathrm{s,in}}$  & 1\,$\mu$W  & 1\,$\mu$W\\
$L$   & 340\,m & 1011\,m\\
$\gamma$ & 12\,W$^{-1}$.km$^{-1}$ & 11.3\,W$^{-1}$.km$^{-1}$\\
$\alpha$   & 0\,dB/km  & 0\,dB/km\\
$D_{\lambda}$  & 0.02\,ps.km$^{-1}$.nm$^{-2}$ & 0.017\,ps.km$^{-1}$.nm$^{-2}$\\
$\delta\lambda_{\mathrm{ofs}}$ &  $-2\,\mathrm{nm}$ &  $-2\,\mathrm{nm}$\\
\hline
\end{tabular}
\caption{Values of the parameters used in the plots of Fig.\,\ref{Figure03}. $P_{j,\mathrm{in}}$: input power of wave labeled $j$ in Fig.\,\ref{Figure02}; $L$: fiber length; $\gamma$: nonlinear coefficient; $\alpha$: fiber attenuation; $D_{\lambda}$: dispersion slope; $\delta\lambda_{\mathrm{ofs}}$: wavelength offset of the signal  with respect to the fiber zero-dispersion wavelength $\lambda_{ZDW}$. The propagation constant $\beta(\omega_{j})$ is expanded at $2^{\mathrm{nd}}$ order around the signal frequency $\omega_{s}$.}
\label{table:1}
\end{table}

\begin{figure}[h]
\centering
\includegraphics[width=1\linewidth]{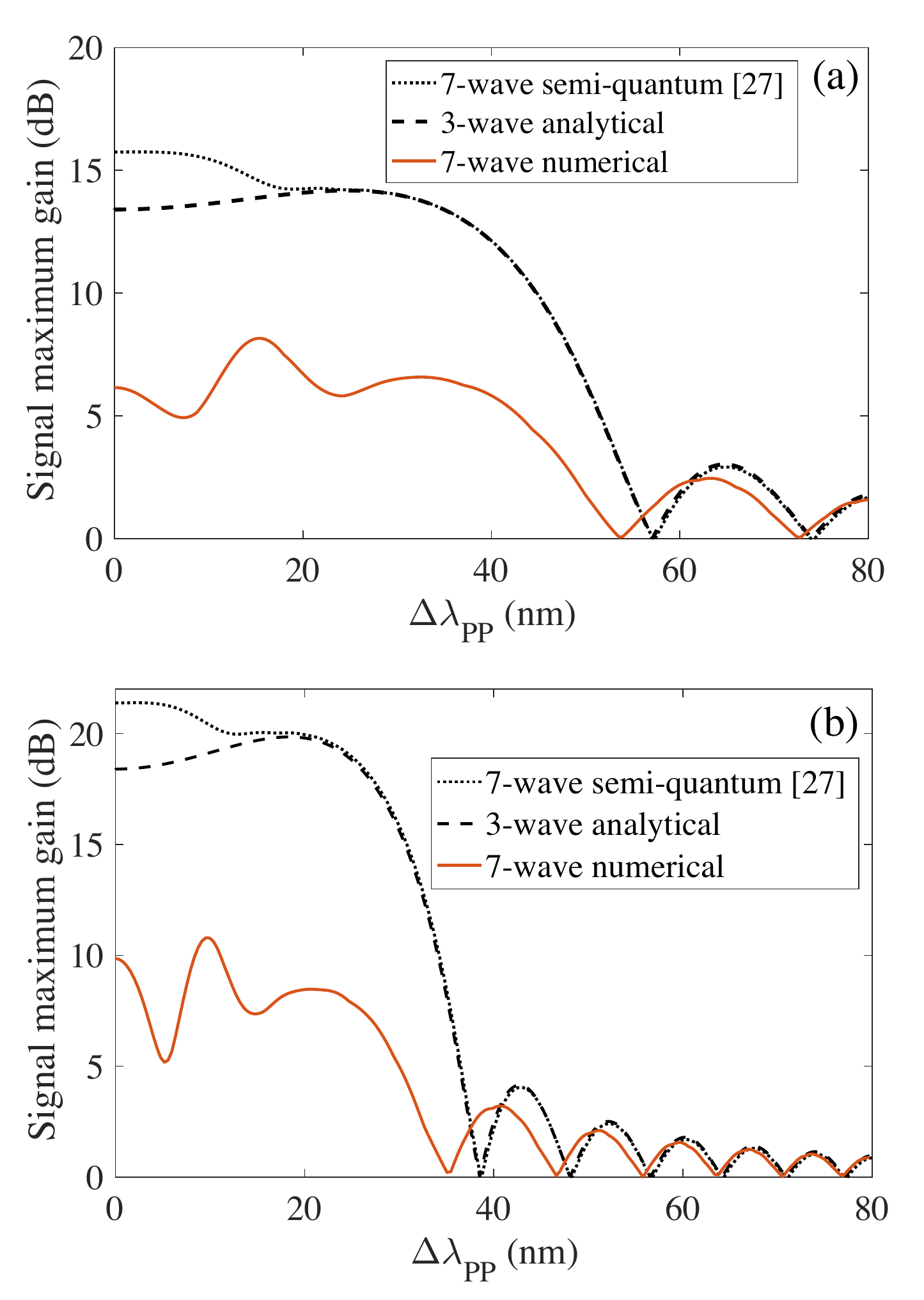}
\caption{Maximum signal gain spectra versus pump-pump wavelength separation $\Delta\lambda_{\mathrm{PP}}$ for two different sets of parameters given in Table\,\ref{table:1}. Dashed black line: 3-wave model; full red line: 7-wave numerical model; dotted black line: 7-wave semi-quantum model of Ref.\,\cite{inoue2019influence}}\label{Figure03}
\end{figure}

The two sets of parameters labeled (a) and (b) in Table \ref{table:1} correspond respectively to the ones used by Inoue in Ref.\,\cite{inoue2019influence} (except the value of $\delta\lambda_{\mathrm{ofs}}$, which is fixed in the present example) and the ones of Ref.\,\cite{xie2015investigation}. In this latter case the fiber parameters are those of one of our experiments \cite{Labidi2018}. These two sets of parameters are used to compute the PSA gain spectra of Figs.\,\ref{Figure03}(a) and \ref{Figure03}(b), respectively. In each of these plots, the PSA maximum gain, obtained by choosing for each set of parameter values the relative phase that maximizes the signal gain, is plotted for the three models as a function of the pump-pump separation $\Delta\lambda_{\mathrm{PP}}$. One can see that in both cases, when $\Delta\lambda_{\mathrm{PP}}$ becomes large, the three models end up giving the same results. But this situation is not very interesting because it corresponds to small gains in a region where the phase mismatch is important. On the contrary, when $\Delta\lambda_{\mathrm{PP}}$ is small and the gain is larger, we can observe strong discrepancies between the three models. In particular, as already stressed in Ref.\,\cite{xie2015investigation}, the numerical 7-wave model shows a very strong degradation of the gain with respect to the 3-wave model. This feature has been shown to be related to the emergence of strong extra tones beyond the two pumps and the signal, such as those labeled 4-7 in Fig.\,\ref{Figure02} \cite{xie2015investigation}. Moreover, in this region and for the two sets of parameters, we can see that the 7-wave semi-quantum model (dotted black line in Fig.\,\ref{Figure03}) gives predictions very close to the ones of the 3-wave model. This shows that the 7-wave semi-quantum model is not valid in this region. Since the values of the gain calculated by this model are not reliable, we thus conclude that we cannot rely on the predictions of this model to calculate the PSA noise figure in the parameter regions where the 3-wave model fails. However, this region had led to the prediction of possible gain enhancements permitted by the extra interactions involving the high-order pumps and high-order idlers \cite{xie2015investigation}. We thus develop in the following a semi-classical noise model based on the numerical 7-wave model in order to calculate the NF of the PSA in such situations.
\section{Semi-classical treatment of the noise figure} \label{Sec3}
In this section, we evaluate the noise characteristics of a dual-pump PSA using a semi-classical approach. Following Ref.\,\cite{tong2012ultralow}, this approach consists in describing the interacting waves classically with their complex amplitudes, while the quantum noise contributions are taken as an additive Gaussian noises that have a zero mean value and a variance of half photon energy. This quantity $h\nu/2$ is interpreted as being the minimum value for the quantized electromagnetic field energy of an harmonic oscillator and is called zero-point energy or vacuum fluctuation energy \cite{loudon2000quantum}. As a consequence of the presence of such a zero-point energy, it is shown that for a PIA with gain $G_{\mathrm{PIA}}$, and in order to respect the Heisenberg inequalities, the minimum output noise power, or the amplified spontaneous emission (ASE) power generated by a PIA, is $ P_{\mathrm{N}} = h\nu B_{0}(G_{\mathrm{PIA}}-1)/2$. It corresponds to the amplification of the vacuum noises power $\delta P = h\nu B_{0}/2$ present at the amplifier input, where $B_{0}$ is the bandwidth of the detection \cite{desurvire2002erbium, heffner1962fundamental}. In the present paper, the extra quantum noise falling into the signal due to its interaction with the high-order idlers and pumps, which are fed with vacuum fluctuations only at the input of the fiber, is treated by including a vacuum noise power at the input of these high-order waves. The interaction between these incident vacuum noises and the pumps and signal along the fiber, and its impact on the output signal power and noise, is evaluated by carrying out numerical simulations of the seven-wave coupled equations. By solving the whole set of equations, and by calculating semi-classically the intensity noise of the signal, a value of the noise figure is deduced.

\subsection{Impact of high-order waves on signal power evolution}\label{Sec3A}
As discussed previously, the presence of high-order waves gives rise to an extra noise, generated from the coupling of their input vacuum fluctuations. To better understand the impact of these high-order waves on the signal power during the amplification process, we plot the evolution of the output signal power versus the fiber length, with and without adding a small amount of power, mimicking vacuum noise, at the input of high-order waves.

In the following of the paper, all simulations are performed with the values of the  parameters given in Table \ref{table:2}. The propagation constant $\beta$ is expanded at  $4^{th}$ order around the signal frequency $\omega_{s}$.
\begin{table}[h]
\centering
\begin{tabular}{l c}
\hline
Parameters &  Values   \\
\hline
Pump 1 power $P_{1,\mathrm{in}}$   & 0.1\,W \\
Pump 2 power $P_{2,\mathrm{in}}$   & 0.1\,W \\
Signal power $P_{\mathrm{s,\mathrm{in}}}$    & 1\,$\mu$W\\
$L$    & 1011\,m\\
$\gamma$  & 11.3\,W$^{-1}$.km$^{-1}$\\
$\alpha$   & 0.9\,dB/km\\
$D_{\lambda}$  & 0.017\,ps.km$^{-1}$.nm$^{-2}$ \\
\hline
\end{tabular}
\caption{Values of the parameters used in the plots of Section \ref{Sec3}. These parameters are similar to the ones of the second column of Table \ref{table:1}, except the non zero fiber attenuation.}
\label{table:2}
\end{table}

\begin{figure}[htbp]
\centering
\includegraphics[width=1\columnwidth]{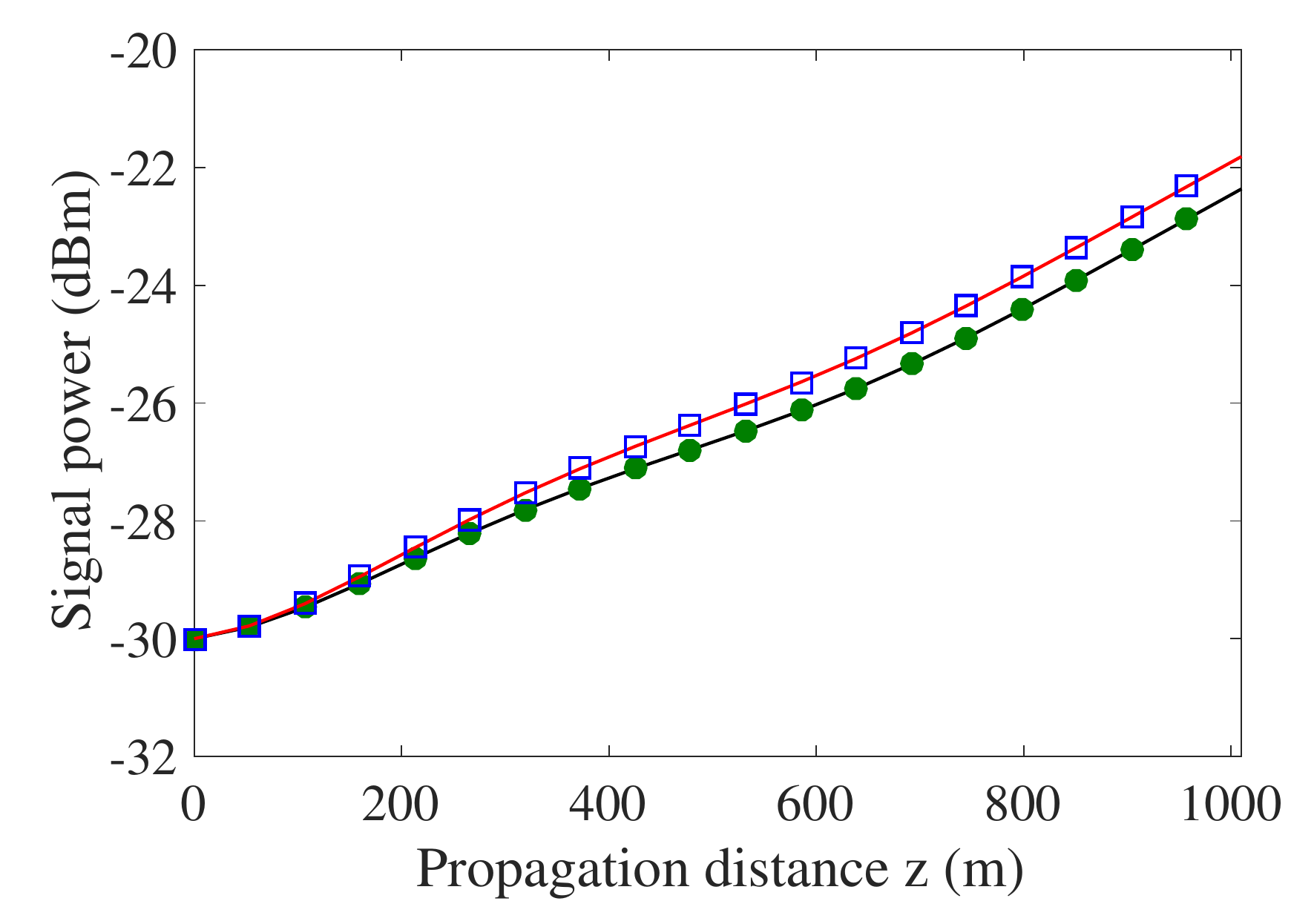}
 \caption{Signal power evolution versus fiber length $z$. Full black line: input powers $P_{j,\mathrm{in}}=0$ for $j=4..7$. Filled circles: $P_{4,\mathrm{in}}=P_{5,\mathrm{in}}=1\,\mathrm{nW}$ and $P_{6,\mathrm{in}}=P_{7,\mathrm{in}}=0$. Full red line: $P_{4,\mathrm{in}}=P_{5,\mathrm{in}}=0$ and $P_{6,\mathrm{in}}=P_{7,\mathrm{in}}=1\,\mathrm{nW}$. Open squares: $P_{j,\mathrm{in}}=1\,\mathrm{nW}$ for $j=4..7$. The other parameters are given in Table \ref{table:2} with $\delta\lambda_{\mathrm{ofs}} = 0$ and a separation between  the pumps $\Delta\lambda_{\mathrm{PP}}=0.32\,\mathrm{nm}$.}\label{Figure04}
\end{figure}
Figure \ref{Figure04} shows that the output signal power is much more sensitive to the injection of a small input power (1~nW in the case of this figure) in the  high-order idlers than in the high-order pumps. This result can be attributed to the fact that the high-order idlers $A_{6}$ and $A_{7}$ (see Fig.\,\ref{Figure02}) are directly coupled to the signal through the FWM processes involving the pumps, while the high-order pumps $A_{4}$ and $A_{5}$ are not directly coupled to the signal (see Eq. \ref{Eq16}).

Consequently, in the following, we neglect the contribution of the high-order pumps 4 and 5 to the signal noise and we consider only the contribution of the vacuum  fluctuations entering the modes of the high-order idlers labeled 6 and 7 in  Fig.\,\ref{Figure02}.

\subsection{Calculation of the excess noise induced by the high-order idlers}\label{Sec3B}
We thus consider the 7-wave situation of Fig.\,\ref{Figure02}, and, following the semi-classical approach of Ref.\,\cite{tong2012ultralow}, we model the  signal amplitude at the output of the PSA of gain $G$ as:
\begin{equation}\label{Eq19}
A_{\mathrm{s,out}} = \sqrt{G}(A_{\mathrm{s,in}} + \delta A_{\mathrm{s,in}}) + \alpha_6\  \delta A_{6,\mathrm{in}} + \alpha_7\ \delta A_{7,\mathrm{in}}\ ,
\end{equation}
where, in agreement with the discussion of Section \ref{Sec3A} and Fig.\,\ref{Figure04}, the coefficients $\alpha_6$ and $\alpha_7$ hold for the transfer of the input vacuum fluctuations $\delta A_{6,\mathrm{in}}$ and $\delta A_{7,\mathrm{in}}$ injected in the high-order idlers to the signal.
In Eq.\,(\ref{Eq19}), we suppose that all field amplitudes, such as $A_{\mathrm{s,out}}$ or $A_{\mathrm{s,in}}$, are normalized in such a way that their square modulus has the units of an energy, i. e., a power per Hertz. We take $A_{\mathrm{s,in}}$ as real, and the input fluctuations $\delta P_i$ are treated classically. For an input coherent state or standard input vacuum fluctuations, we have:
\begin{align}
\langle \delta A_j\rangle&=0\,\label{Eq20}\\
\langle \delta A_j^2\rangle&=0\,\label{Eq21}\\
\langle |\delta A_j|^2\rangle&=\frac{h\nu_{j}}{2}\,.\label{Eq22}
\end{align}
After detection with a perfect photodetector in a bandwidth $B$, the output signal leads to the following photocurrent:
\begin{eqnarray}
I_{\mathrm{s,out}}&=&R_0 B\left|A_{\mathrm{s,out}}\right|^2\nonumber\\
&=&R_0 B\left\{GA_{\mathrm{s,in}}^2+GA_{\mathrm{s,in}}(\delta A_{\mathrm{s,in}}+\delta A_{\mathrm{s,in}}^*)\right.\nonumber\\&&+\sqrt{G}A_{\mathrm{s,in}}(\alpha_6\delta A_{6,\mathrm{in}}+\alpha_6^*\delta A_{6,\mathrm{in}}^*)\nonumber\\&&\left.+\sqrt{G}A_{\mathrm{s,in}}(\alpha_7\delta A_{7,\mathrm{in}}+\alpha_7^*\delta A_{7,\mathrm{in}}^*)\right\}\ ,\label{Eq24}
\end{eqnarray}
with $R_0=e/h\nu$. The average value and the variance of this photocurrent are found to be given by
\begin{eqnarray}
    \langle I_{\mathrm{s,out}}\rangle&=&R_0 B G A_{\mathrm{s,in}}^2\,,\\
    \langle \Delta I_{\mathrm{s,out}}^2\rangle&=&2R_0^2B^2A_{\mathrm{s,in}}^2\left[G^2\langle |\delta A_{\mathrm{s,in}}|^2\rangle\right.\nonumber\\&&\left.+G|\alpha_6|^2\langle |\delta A_{6,\mathrm{in}}|^2\rangle+G|\alpha_7|^2\langle |\delta A_{7,\mathrm{in}}|^2\rangle\right]\ 
\end{eqnarray}
Similarly, the photocurrent that would be  obtained by detecting the input signal would have the following average value and variance:
\begin{eqnarray}
    \langle I_{\mathrm{s,in}}\rangle&=&R_0 B  A_{\mathrm{s,in}}^2\,,\\
    \langle \Delta I_{\mathrm{s,in}}^2\rangle&=&2R_0^2B^2A_{\mathrm{s,in}}^2\langle |\delta A_{\mathrm{s,in}}|^2\rangle .
\end{eqnarray}
The contribution of the high-order idlers labeled 6 and 7 in Fig.\,\ref{Figure02} to the PSA noise figure is thus given by:
\begin{eqnarray}
NF_{6,7}&=&\frac{\langle I_{\mathrm{s,in}}\rangle^2/\langle \Delta I_{\mathrm{s,in}}^2\rangle}{\langle I_{\mathrm{s,out}}\rangle^2/\langle \Delta I_{\mathrm{s,out}}^2\rangle}\nonumber\\
&=&1+\frac{1}{G} \frac{|\alpha_6|^2\langle |\delta A_{6,\mathrm{in}}|^2\rangle+|\alpha_7|^2\langle |\delta A_{7,\mathrm{in}}|^2\rangle}{\langle |\delta A_{\mathrm{s,in}}|^2\rangle}\ .\label{Eq28}
\end{eqnarray}
Using Eq.\,(\ref{Eq22}), Eq.\,(\ref{Eq28}) finally becomes:
\begin{equation}
NF_{6,7}=1+\frac{|\alpha_6|^2+|\alpha_7|^2}{G}\ .\label{Eq29}
\end{equation}

\begin{figure}[htbp]
\centering
\includegraphics[width=0.9\columnwidth]{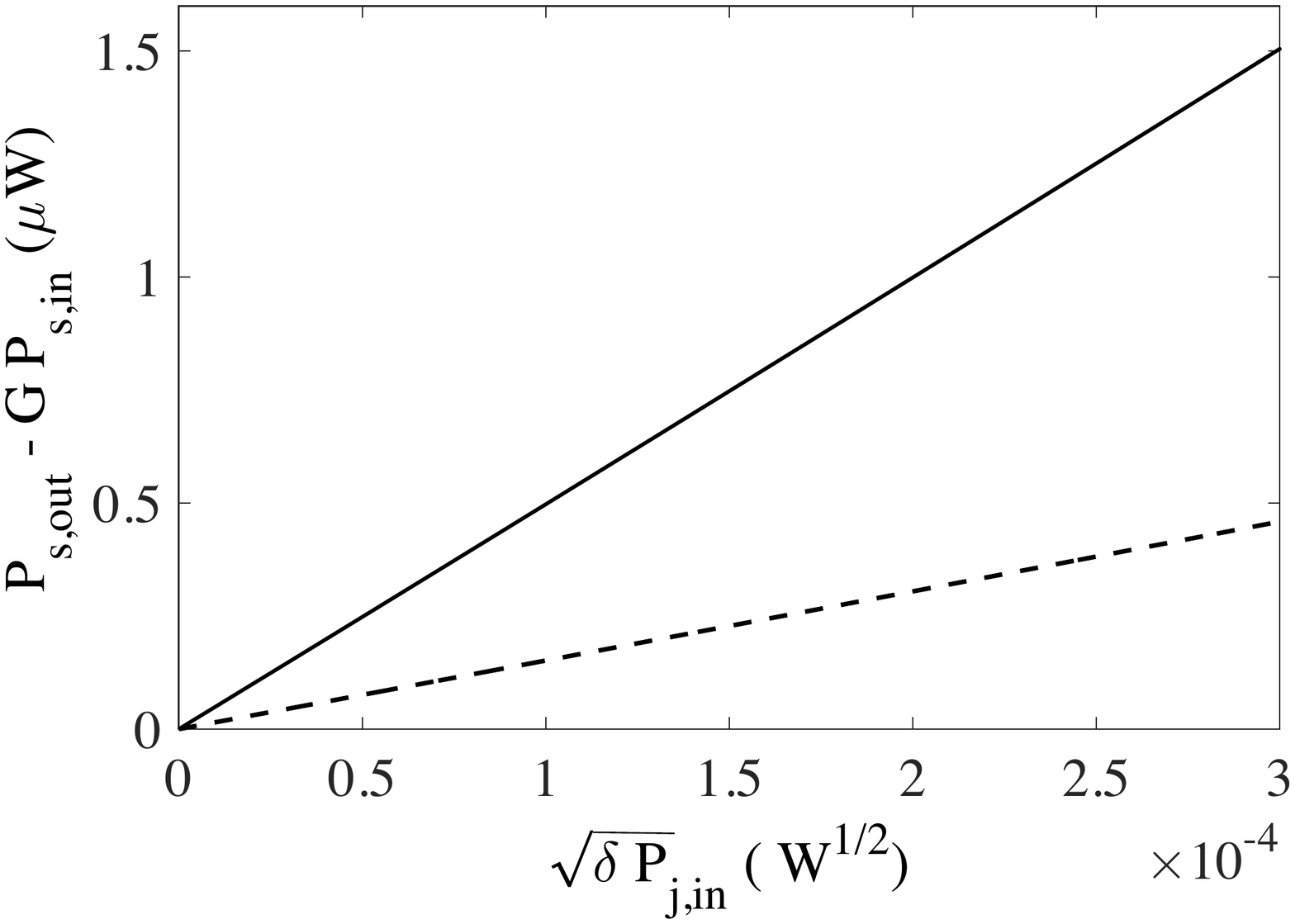}
 \caption{Signal output power increase versus  $\sqrt{\delta P_{6,\mathrm{in}}}$ (dashed line) and $\sqrt{\delta P_{7,\mathrm{in}}}$ (full line). Same parameters as in Fig.\,\ref{Figure04} except $\Delta\lambda_{\mathrm{PP}}=28\,\mathrm{nm}$.}\label{Figure05}
\end{figure}

In order to  determine $|\alpha_6|^2$ and $|\alpha_7|^2$, we notice that the result of Eq.\,(\ref{Eq28}) is based on Eq.\,(\ref{Eq19}), which supposes that the signal power at the output of the PSA is proportional to the square root of the high-order idler input powers. Indeed, if one injects small powers $P_{6,\mathrm{in}}$ and $P_{7,\mathrm{in}}$ in high-order idler modes 6 and 7 at the input of the fiber, Eq.\,(\ref{Eq19}) leads, at the limit where $P_{6,\mathrm{in}},P_{7,\mathrm{in}}\ll P_{\mathrm{s,in}}$, to:
\begin{equation}
P_{\mathrm{s,out}}=GP_{\mathrm{s,in}}+2\sqrt{G}\sqrt{P_{\mathrm{s,in}}}\left(|\alpha_6|\sqrt{P_{6,\mathrm{in}}}+|\alpha_7|\sqrt{P_{7,\mathrm{in}}}\right)\ .\label{Eq30}
\end{equation}
This behavior can be checked by simulating the evolution of $P_{\mathrm{s,out}}$ as a function of $P_{6,\mathrm{in}}$ and $P_{7,\mathrm{in}}$. The result of such simulations is reproduced in Fig.\,\ref{Figure05}, which was obtained with the same parameters as those used to obtain Fig.\,\ref{Figure04} and for the value of the relative phase that maximizes the signal gain. This figure reproduces the evolution of the increase in signal output power $P_{\mathrm{s,out}}-GP_{\mathrm{s,in}}$ as a function of $\sqrt{P_{6,\mathrm{in}}}$ and $\sqrt{P_{7,\mathrm{in}}}$.

 This figure confirms that $P_{\mathrm{s,out}}$ evolves linearly with $\sqrt{P_{6,\mathrm{in}}}$ and $\sqrt{P_{7,\mathrm{in}}}$, as expected from Eq.\,(\ref{Eq30}). We can thus take the slopes $\partial P_{\mathrm{s,out}}/\partial\sqrt{P_{6,\mathrm{in}}}$ and $\partial P_{\mathrm{s,out}}/\partial\sqrt{P_{7,\mathrm{in}}}$ as constants. This permits to rewrite  Eq.\,(\ref{Eq29}) in the following form:
\begin{equation}
NF_{6,7}=1 + \frac{1}{4G^2P_{\mathrm{s,in}}}\left[\left(\frac{\partial P_{\mathrm{s,out}}}{\partial\sqrt{P_{6,\mathrm{in}}}}\right)^2+\left(\frac{\partial P_{\mathrm{s,out}}}{\partial\sqrt{P_{7,\mathrm{in}}}}\right)^2\right]\ \label{Eq32}
\end{equation}
Finally combining this expression with Eq.\,(\ref{Eq14N1}) leads to the following expression:
\begin{equation}
NF=1+\frac{G_{min}}{G_{max}} + \frac{1}{4G^2P_{\mathrm{s,in}}}\left[\left(\frac{\partial P_{\mathrm{s,out}}}{\partial\sqrt{P_{6,\mathrm{in}}}}\right)^2+\left(\frac{\partial P_{\mathrm{s,out}}}{\partial\sqrt{P_{7,\mathrm{in}}}}\right)^2\right]\,. \label{Eq33}
\end{equation}
This expression permits to deduce the value of the noise figure from the slopes of the plots like those of Fig.\,\ref{Figure05}, which are easily obtained from the simulations.

\section{Application and discussion}\label{Sec4}
We now apply the formalism developed in the preceding section to the calculation of the noise figure in two different situations. Indeed, depending on the dispersion characteristics of the fiber and the spectral distribution of the waves, it has been shown \cite{xie2015investigation} that the emergence of the high-order pumps and idlers can lead either to a decrease or an increase of the maximum signal gain compared to what is expected from the three-wave model. We thus investigate in the two following subsections whether the interplay with the high-order idlers and pumps gives rise to a degradation of the noise figure in these two situations.
\subsection{Case where the high-order idlers decrease the gain} 
The situation considered here corresponds to  the values of the parameters summarized in Table\,\ref{table:2}. The wavelength of the signal coincides in this case with the zero-dispersion wavelength: $\delta\lambda_{\mathrm{ofs}}=0$. 
\begin{figure}[htbp]
\centering
\includegraphics[width=0.8\columnwidth]{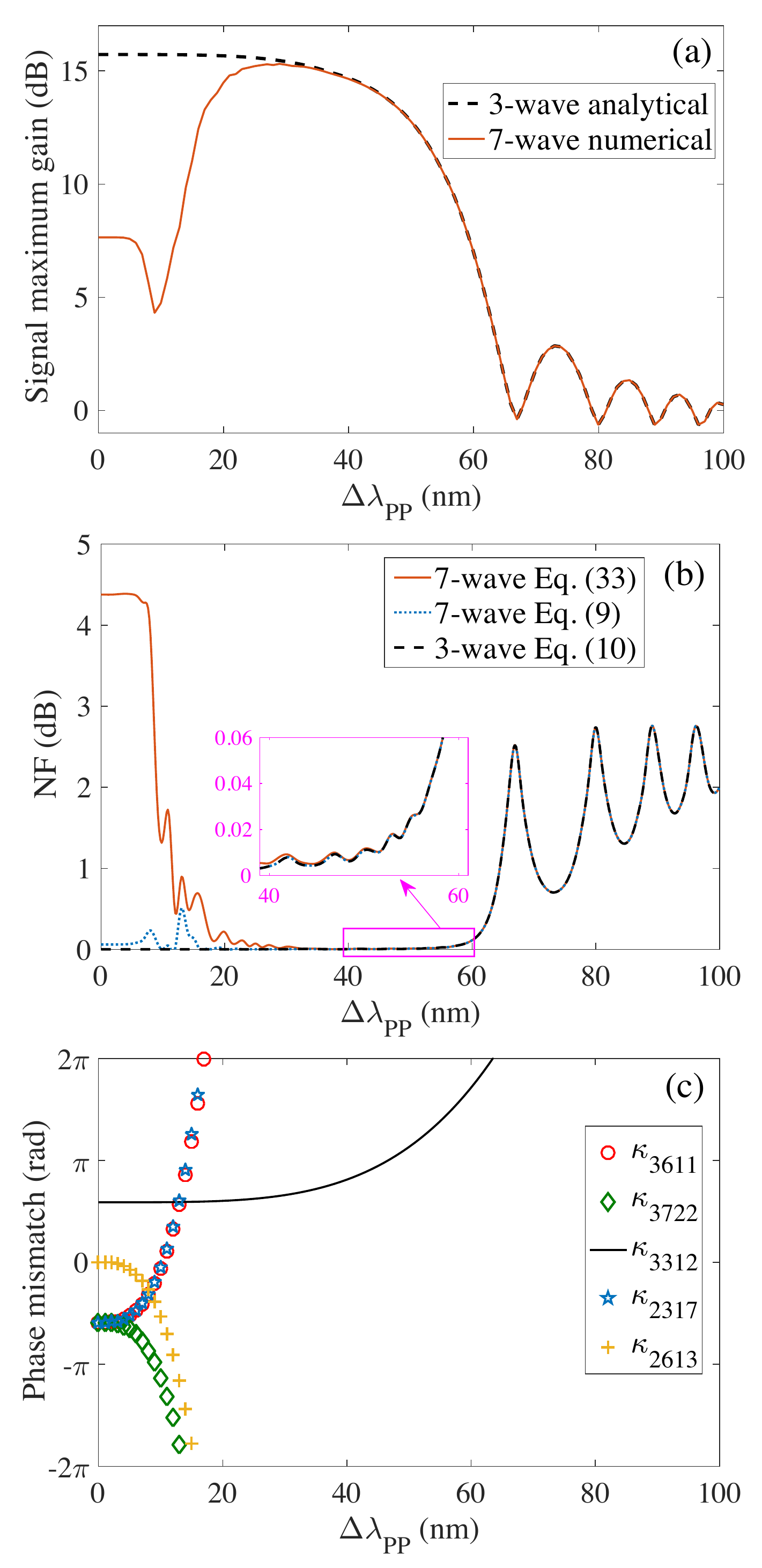}
\caption{Case where $\delta\lambda_{\mathrm{ofs}}=0$. (a) Signal maximum gain versus pump-pump wavelength separation $\Delta\lambda_{\mathrm{PP}}$  for the 3-wave (dashed black line) and 7-wave (full red line) models. (b) Signal noise figure versus $\Delta\lambda_{\mathrm{PP}}$ based on three different expressions (see text), with the inset showing a zoom on the $40\,\mathrm{nm}\leq\Delta\lambda_{\mathrm{PP}}\leq60\,\mathrm{nm}$ region. (c) Phase mismatch  of several FWM processes versus $\Delta\lambda_{\mathrm{PP}}$.}\label{Figure06}
\end{figure}
Figure \ref{Figure06}(a) shows a comparison between the maximum gain of the PSA according to the 3-wave model (Equation \ref{Eq6}, dashed line) and the numerical 7-wave model (full line), as a function of the separation $\Delta\lambda_{\mathrm{PP}}$ between the two pumps. As already observed in Ref.\,\cite{xie2015investigation}, the 7-wave model predicts a strong reduction of the gain for values of $\Delta\lambda_{\mathrm{PP}}$ smaller than 30\,nm. Since this gain decrease is associated with the emergence of the high-order idlers and pumps, we expect the model of Section \ref{Sec3}.\ref{Sec3B} to lead to a degradation of the NF of the PSA. This is indeed  what is observed in Figure \ref{Figure06}(b), which compares three different expressions for the noise figure at the maximum gain of the PSA: i) the one based on the three-wave  model (Eq. \ref{Eq15}, dashed black line); ii) the one that just takes into account the values of the minimum and maximum gains predicted by the 7-wave model and injects them into Eq. (\ref{Eq14N1}) (dotted blue line); iii) the one  based on Eq.\,(\ref{Eq33}) that takes into account the transfer to the signal of the  vacuum fluctuations injected in the high-order idlers (full red line). 

One can see that when the two models predict the same gain, i.e., for $\Delta\lambda_{\mathrm{PP}}>30\,\mathrm{nm}$ (see Fig. \ref{Figure06}(a)), the three expressions converge to the same value of the NF. This is consistent with the  fact that the high-order idlers and pumps are negligible in this parameter region. However, for small values of $\Delta\lambda_{\mathrm{PP}}\lesssim 10\,\mathrm{nm}$, Eq.\,(\ref{Eq33}) predicts a strong  increase of the noise figure (up to 4.5\,dB) with respect to the two other expressions. This increase of the noise figure comes from the high-order idlers, as confirmed by the plots of Fig.\,\ref{Figure06}(c). This figure reproduces the evolution versus $\Delta\lambda_{\mathrm{PP}}$ of the total phase  mismatches of the  different four-wave mixing processes that participate to the output signal noise. These phase mismatches are given by $\kappa_{mnkl}L$, where the coefficients $\kappa_{mnkl}$ are calculated using Eq.\,(\ref{Eq17}) at the output of the fiber. We have selected only the processes present in the equation of evolution of the signal (Eq.\,\ref{Eq16}) that involve the two pumps, one of the high-order idlers, and the signal. Indeed, the processes that involve only one of the pumps or no pump at all are much weaker than the ones we take into account. Moreover, we have seen in Section \ref{Sec3A} (see Fig.\,\ref{Figure04}) that the high-order pumps do not significantly contribute to the output signal noise. 

The plots of Fig.\,\ref{Figure06}(c) confirm that the  processes that contribute to the increase of the signal power are those that involve the high-order idlers (waves labeled 6 and 7 in Fig.\,(\ref{Figure02})). Indeed, one can see that the bandwidth ($\Delta\lambda_{\mathrm{PP}}\lesssim 10\,\mathrm{nm}$) of the increase of the noise figure in Fig.\,\ref{Figure06}(b) corresponds to the domain in which the phase mismatch coefficients for the FWM processes involving the high-order idlers are in the interval $[-\pi,\pi]$, showing that these processes are efficient. As soon as those processes are no longer phase matched, i. e., $|\kappa_{mnkl}L|>\pi$, the noises of the high-order idlers stop contaminating the signal and the noise figure retrieves its values predicted by the three-wave model.

\subsection{Case where the high-order idlers increase the gain}
\begin{figure}[htbp]
\centering
\includegraphics[width=0.8\columnwidth]{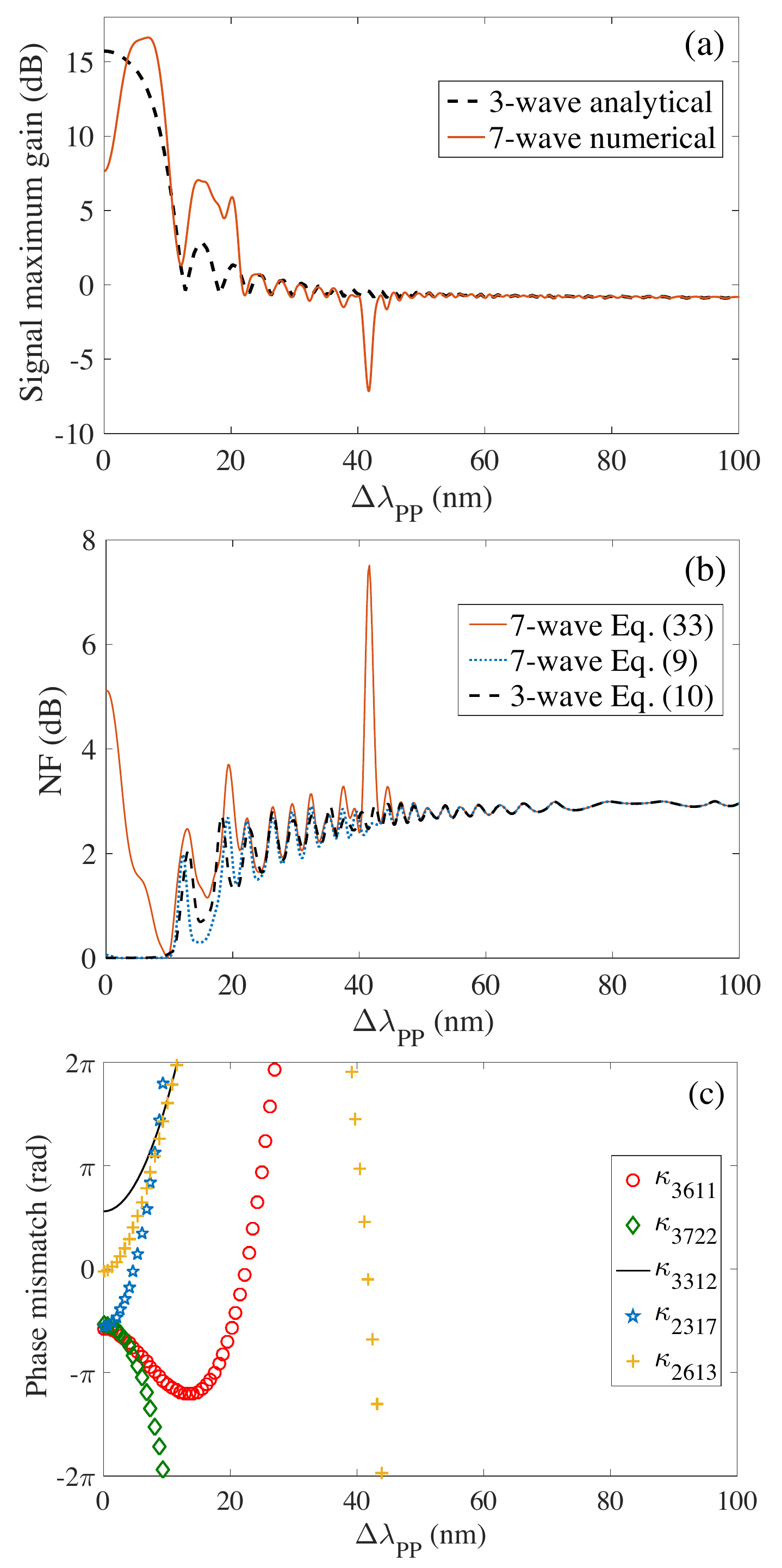}
\caption{(a-c) Same as Fig.\,\ref{Figure06} for $\delta\lambda_{\mathrm{ofs}} = 10\,  \mathrm{nm}$}\label{Figure07}
\end{figure}

We now turn to a situation in which the interplay between the high-order idlers and pumps and the signal is a priori favorable, in the sense that it permits to increase the maximum signal gain. As can be seen in Fig.\,(\ref{Figure07}), this can be obtained by  shifting the wavelengths of the signal and the pump by $\delta\lambda_{\mathrm{ofs}} = 10\,  \mathrm{nm}$ with respect to the zero-dispersion wavelength. Then, for $\Delta\lambda_{\mathrm{PP}}=7.1\,\mathrm{nm}$, the 7-wave model predicts a maximum gain equal to 16.6\,dB, larger than the gain predicted by the 3-wave model for this pump-pump separation (12.8\,dB), and even larger than the 15.7\,dB gain predicted by the 3-wave model for $\Delta\lambda_{\mathrm{PP}}=0$.

The corresponding evolution of the noise figure is plotted in Fig.\,\ref{Figure07}(b), according to the same three expressions as in Fig.\,\ref{Figure06}(b). Here also, a signal excess noise is observed for small values of $\Delta\lambda_{\mathrm{PP}}$ when one takes into account the noise transfer from the high-order idlers (full red line in Fig.\,\ref{Figure07}(b)). Figure\,\ref{Figure07}(c) confirms that this noise increase is due to the transfer of the noise from the high-order idlers, because the bandwidth of this noise increase $\Delta\lambda_{\mathrm{PP}}\lesssim 10\,\mathrm{nm}$ corresponds to the range in which the corresponding FWM processes are phase matched ($|\kappa_{mnkl}L|<\pi$). 

However, here, contrary to the case $\delta\lambda_{\mathrm{ofs}} =0$ of Fig.\,(\ref{Figure06}), two of these processes get phase matched again for larger values of $\Delta\lambda_{\mathrm{PP}}$ thanks to the fact that the linear and nonlinear parts of their phase mismatches have opposite signs. This is the case of the process involving the pump 1 with the signal and the high-order idler 6 ($\kappa_{3611}$ in Fig.\,\ref{Figure07}(c)), which gets perfectly phase matched again for $\Delta\lambda_{\mathrm{PP}}\simeq17.5\,\mathrm{nm}$, leading to another increase of the noise figure according to the 7-wave model compared with the 3-wave model. 

The same type of phenomenon occurs around $\Delta\lambda_{\mathrm{PP}}\simeq41.5\,\mathrm{nm}$. Around this value of the pump-pump separation, the gain predicted by the 7-wave model exhibits a strong narrow dip (see the red full line in Fig.\,\ref{Figure07}(a)), which is accompanied by a strong increase of the noise figure (see Fig.\,\ref{Figure07}(b)). Figure\,\ref{Figure07}(c) shows that this phenomenon is due to the fact that the FWM process involving the high-order idler labeled 6 with the two pumps and the signal ($\kappa_{2613}$ in Fig.\,\ref{Figure07}(c)) gets phase matched again around this value of $\Delta\lambda_{\mathrm{PP}}$, leading to an efficient noise transfer from idler number 6 to the signal.

Coming back to the behavior of the PSA according to the 7-wave model in the vicinity of the gain maximum, i. e., around $\Delta\lambda_{\mathrm{PP}}=7.1\,\mathrm{nm}$, the detailed observation of Figs.\,\ref{Figure07}(a) and \ref{Figure07}(b) permits to conclude that this situation is a good trade-off between the gain increase and the NF degradation due to the high-order idlers. Indeed, the few dB gain increase is accompanied by a relatively modest degradation of the noise figure (equal to 0.95\,dB for $\Delta\lambda_{\mathrm{PP}}=7.1\,\mathrm{nm}$), which remains much below the 3-dB limit encountered in the case of a PIA. The presence of the high-order idlers can thus be helpful for the gain without being too detrimental to the noise figure.

\section{Conclusion}
In conclusion, the noise performance of degenerate dual-pump phase sensitive amplifier was investigated thanks to a semi-classical approach based on a 7-wave model.  This approach was adopted after having observed that other approaches based on analytical calculations did not always predict the correct value for the gain in the range of parameters we are interested in. 

In our approach, no assumptions were made on which FWM processes among the 7 waves should be neglected, as compared with the semi-quantum approach. Numerical simulations of the 7-wave coupled equations were carried out, leading to accurate results for the noise figure. 

In the range of parameters we considered here, compared with the standard 3-wave PSA, we have seen that the high-order idlers can degrade the noise figure of the amplifier when the four-wave mixing processes that transfer the vacuum fluctuations injected in these modes into the signal become efficient. Conversely, the presence of the high order pumps does not lead to any significant degradation of the signal noise. For stronger nonlinearities (longer fiber, larger value of the nonlinear coefficient, and/or higher pump powers), more general models involving more waves and more nonlinear processes should be considered \cite{Qian2017}, at the cost of an severely increased complexity.

In spite of this degradation of the noise figure, we have seen that by adjusting the dispersion of the fiber and the frequency distribution of the signal and pumps, one can take advantage of the emergence of the extra waves to enhance the signal gain, while maintaining the noise figure of the amplifier below 1~dB. This interesting results opens interesting perspectives of application and is promising in view of optimizing the noise in PSA systems. For example, this could be done by coupling our model with a genetic algorithm, as was already performed to optimize the gain of a PSA \cite{Li2017}.

\section*{Disclosures}
The authors declare no conflicts of interest.


\bibliography{sample}

\end{document}